\documentclass[conference]{IEEEtran}
\IEEEoverridecommandlockouts

\usepackage[letterpaper, left=0.625in, right=0.625in, top=0.75in, bottom=1in]{geometry}

\IEEEaftertitletext{\vspace{-2\baselineskip}}

\usepackage{balance}
\usepackage{cite}
\usepackage{amsmath,amssymb,amsfonts}
\usepackage{algorithmic}
\usepackage{graphicx}
\usepackage{textcomp}
\usepackage{xcolor}
\usepackage{pgfplots}
\pgfplotsset{compat=1.17}
\usepackage{booktabs}
\usepackage{float}
\usepackage{bm}
\def\BibTeX{{\rm B\kern-.05em{\sc i\kern-.025em b}\kern-.08em
    T\kern-.1667em\lower.7ex\hbox{E}\kern-.125emX}}
    
\begin{document}

\title{A Hodge-FAST Framework for High-Resolution Dynamic Functional Connectivity\\
Analysis of Higher Order Interactions in EEG Signals}

\author{
\IEEEauthorblockN{
    Om Roy\textsuperscript{1,*}, Yashar Moshfeghi\textsuperscript{1}, Jason Smith\textsuperscript{3}, 
    Agustin Ibanez\textsuperscript{4,5},\\ 
    Mario A. Parra\textsuperscript{2}, Keith M. Smith\textsuperscript{1}
}
\IEEEauthorblockA{
    \textsuperscript{1}Computer and Information Sciences, University of Strathclyde, Glasgow, UK \\
    \textsuperscript{2}Psychological Sciences and Health, University of Strathclyde, Glasgow, UK \\
    \textsuperscript{3}Department of Mathematics, Nottingham Trent University, Nottingham, UK \\
    \textsuperscript{4}Latin American Brain Health Institute (BrainLat), Universidad Adolfo Ibañez, Santiago, Chile \\
    \textsuperscript{5}Global Brain Health Institute, Trinity College Dublin, Dublin, Ireland
}
\thanks{*Corresponding Author: Om Roy, o.roy.2022@uni.strath.ac.uk}
}

\maketitle

\begin{abstract}
We introduce a novel framework that integrates Hodge decomposition with Filtered Average Short-Term (FAST) functional connectivity to analyze dynamic functional connectivity (DFC) in EEG signals. This method leverages graph-based topology and simplicial analysis to explore transient connectivity patterns at multiple scales, addressing noise, sparsity, and computational efficiency. The temporal EEG data are first sparsified by keeping only the most globally important connections, instantaneous connectivity at these connections is then filtered by global long-term stable correlations. This tensor is then decomposed into three orthogonal components to study signal flows over higher-order structures such as triangle and loop structures. Our analysis of Alzheimer-related MCI patients show significant temporal differences related to higher-order interactions that a pairwise analysis on its own does not implicate. This allows us for the first time to capture higher-dimensional interactions at high temporal resolution in noisy EEG
signal recordings.
 
\end{abstract}

\section{Introduction}

There has been a boom of studies in recent years studying brain recordings processed into networks \cite{Roy2024,Roy2023,hc,sporns1}. Most work predominantly focuses on the use of graph theory, a rich mathematical field that studies networks in the form of pairwise interactions of nodes with accompanying edges \cite{bullmore}. While being very informative and useful in providing network metrics to distinguish between control and diseased states, it neglects the presence of higher-order topological interactions \cite{higher}. The brain is a complex system with various dynamically interacting sub-systems, in order to truly get a thorough understanding of its topology it is crucial to understand its behaviour throughout \textit{time} \cite{DFC,mde}, in particular, how these higher-order interactions reveal complimentary information in addition to traditional pair-wise analysis. 

The electroencephalogram (EEG) is a relatively cheap, non-invasive approach to measure neural activity using electrical currents measured directly on the scalp \cite{hc,alth}. Having a very high temporal resolution \cite{eeg5,adeeg}, unmatched by even more expensive methods such as fMRI, there is a high degree of untapped potential in uncovering useful information throughout time. This potential is majorly impaired by the large amount of inherent noise in the medium, that is, effects of volume conduction are present due to the presence of muscle bone between the recording medium and the source (the neurological components themselves) \cite{mde,Roy2024}. This has motivated the development of noise-robust methods to exploit the temporal resolution of the EEG and in turn determining the statistical correspondence between approximate regions of the brain throughout time, this has been termed as Dynamic Functional Connectivity (DFC) \cite{DFC,kaiser2}.

While there have been many approaches in the estimation of DFC in EEG signals, approaches leveraging methods from Network Science have seen a rise in popularity. In particular, we recently introduced Filtered Average Short-Term (FAST) functional connectivity (FC) \cite{Roy2024}. FAST FC is a novel approach that leverages a stable filter of long-term global correlation computed over all participants involved in a time-locked cognitive task. This is used to filter individual instantaneous connectivity profiles of individual participants. This approach has been shown to be effective even in very noisy conditions while also outperforming traditional spectral approaches such as the Wavelet Transform in DFC analysis of EEG signals. Furthermore, this approach indicates a shift from graph based methods based in the Graph Fourier Domain \cite{gsp} and thus does not require the computationally expensive procedure of eigen-decomposition, making it an interpretable, fast method. 

Methods like FAST FC however, rely solely on pairwise interactions, neglecting the presence of higher order interactions between brain regions. Topological data analysis (TDA) \cite{tda1,top,twins} approaches that utilize persistent homology (PH) \cite{medi,game,topeeg,wang2,ph} use simplicial complexes\cite{simp} to model higher-order interactions. Simplicial complexes generalize the concept of interactions in a network in the sense that 0-way interactions are represented by nodes, 1-way interactions by edges and 2-way interactions are represented by triangles. This can be extended even further to structures such as tetrahedra. Equivalently, the Hodge Laplacian \cite{hodgelap} serves as the generalization of the graph Laplacian to general simplicial complexes. Anand et al. \cite{hodge,hodge2} exploited this relationship to produce the Hodge decomposition of brain networks which broke down pair-wise neurological interactions into gradient, curl and loop flows-- orthogonal components that capture interactions at different orders. The gradient flow captures the behavior of node gradients, the curl flow the behavior of triangle induced flows and the harmonic flows implicates loop-like structures.

The application of the Hodge Decomposition of Brain networks and other topological approaches involving simplicial complexes has been limited to the analysis of static representations of functional connectivity \cite{medi,game,topeeg}. In this work, we introduce a novel framework that integrates Hodge decomposition with FAST FC for analyzing dynamic functional connectivity (DFC) in EEG signals. This approach allows us to extend node-pair interactions to edge flows over triangular complexes, allowing us, for the first time, to capture higher-dimensional interactions at a high temporal resolution in noisy EEG signal recordings. Applying this to two independent cohorts of Sporadic and Familial Alzheimer Disease (AD) related Mild Cognitive Impairment (MCI) patients performing a time-locked shape-colour binding task we find consistent, reproducible significant differences between groups in time intervals that have been previously implicated in the task. This unveils new, potential specific biomarkers for the early detection of the debilitating condition worth further investigation.

\section{Methods}

\subsection{FAST Functional Connectivity}
FAST Functional Connectivity, inspired from the general graph-variate dynamic connectivity framework \cite{gvsa}, provides an ideal platform for the analysis of DFC in EEG signals.

The long-term stable connectivity tensor, \( \bm{C}^{\text{FAST}} \) [1], serves as the foundation for filtering instantaneous edge flows. It is defined as the modulus of the Pearson correlation coefficient, averaged across \( N \) participants:
\begin{equation}
\begin{split}
    C_{ij}^{\text{FAST}} &= \frac{1}{N} \sum_{P=1}^N C_{ij}^{(P)}, \\
    C_{ij}^{(P)} &= 
    \frac{\sum_{t \in T} \big(x_i(t) - \bar{x}_i\big)\big(x_j(t) - \bar{x}_j\big)}%
    {\sqrt{\sum_{t \in T} \big(x_i(t) - \bar{x}_i\big)^2} 
    \sqrt{\sum_{t \in T} \big(x_j(t) - \bar{x}_j\big)^2}}.
\end{split}
\end{equation}

where \( x_i(t) \) is the signal at node \( i \) at time \( t \), and \( \bar{x}_i \) is the temporal mean at node \( i \).

The instantaneous connectivity tensor, \( \bm{\Theta}^{\text{FAST}} \), can be computed using the local Dirichlet energy or the instantaneous correlation. For the local Dirichlet energy, it is given by:
\begin{equation}
    \Theta_{ij}^{\text{FAST}}(t) = 
    \begin{cases} 
        C_{ij}^{\text{FAST}} \cdot \big(x_i(t) - x_j(t)\big)^2, & i \neq j, \\
        0, & i = j.
    \end{cases}
\end{equation}

Alternatively, for the instantaneous correlation, it can be expressed as:
\begin{equation}
    \Theta_{ij}^{\text{FAST}}(t) = 
    \begin{cases} 
        C_{ij}^{\text{FAST}} \cdot \big|(x_i(t) - \bar{x}_i)(x_j(t) - \bar{x}_j)\big|, & i \neq j, \\
        0, & i = j.
    \end{cases}
\end{equation}

Where \( \bar{x}_i \) represents the mean of \( x_i \) over the considered time window.
\subsection{Temporal Hodge Decomposition of FAST Tensor}
In order to ensure computational efficiency when computing higher order structures we sparsify the connectivity tensor \( \bm{\Theta}^{\text{FAST}}(t) \), using the mask matrix \( \bm{M} \), where:

\[
    M_{ij} = 
    \begin{cases} 
    1, & \text{if } C_{ij}^{\text{FAST}} \geq C_K^{\text{FAST}}, \\
    0, & \text{otherwise},
    \end{cases}
\]
where the threshold \( C_K^{\text{FAST}} \) is defined as the value in the connectivity matrix \( C_{ij}^{\text{FAST}} \) above which only the top \( K \)-th percentile of connections are retained, ensuring that only the strongest connections are preserved.

Once the connectivity tensor is sparsified,
\[ \bar{\bm{\Theta}}^{\text{FAST}}(t)  = \bm{M}\odot\bm{\Theta}^{\text{FAST}}(t), \] 

where \(\odot\) is the Hadamard product, we average it over a sliding window of size \( W \) to capture the temporal dynamics while maintaining sparsity:

\[
    \bar{\bm{\Theta}}^{\text{FAST}}(t) = \frac{1}{W} \sum_{t' = t}^{t+W} (\bm{M}\odot \bm{\Theta}^{\text{FAST}}(t')).
\]

Boundary operators are used to capture the interactions between simplicial complexes of different dimensions. For the \( p \)-dimensional boundary operator \( B_p(t) \), we have:

\[
    B_{p_{ij}} =
    \begin{cases}
    1, & \text{if the \( k \)-simplex is part of a \( (k+1) \)-simplex}, \\
    0, & \text{otherwise}.
    \end{cases}
\]

The operator \( B_p \) captures the interactions between lower-dimensional and higher-dimensional simplicial complexes at each time step.

Using these  boundary operators, the \( p \)-th Hodge Laplacian is expressed as:

\[
    \Delta_p = B_p^\top B_p + B_{p+1} B_{p+1}^\top,
\]

where \( B_p^\top B_p \) captures lower-dimensional interactions (e.g., nodes-to-edges), and \( B_{p+1} B_{p+1}^\top \) captures upper-dimensional interactions (e.g., edges-to-triangles).

As described in \cite{hodge, hodge2}, the coboundary operators $\delta_k$ serve as the duals of boundary operators, mapping $k$-simplices to $(k+1)$-simplices. The coboundary operator $\delta_k$ is the adjoint of the boundary operator $\partial_{k+1}$. For real matrices, $\delta_k$ is represented as the transpose of the matrix representation of $\partial_{k+1}$: $\delta_k = B_{k+1}^\top$.

The temporal edge flow \( \bar{\bm{\Theta}}^{\text{FAST}}(t) \) is decomposed into its gradient, curl, and harmonic components via Hodge decomposition:

\[
    \bar{\bm{\Theta}}^{\text{FAST}}(t) = \bar{\bm{\Theta}}_G^{\text{FAST}}(t) + \bar{\bm{\Theta}}_C^{\text{FAST}}(t) + \bar{\bm{\Theta}}_H^{\text{FAST}}(t),
\]

where \( \bar{\bm{\Theta}}_G^{\text{FAST}}(t)\) captures global flows, \( \bar{\bm{\Theta}}_C^{\text{FAST}}(t)\) describes localized cycles (loops), and \( \bar{\bm{\Theta}}_H^{\text{FAST}}(t) \) is the harmonic component, which is divergence-free and curl-free.

The gradient flow, which is influenced by the gradients at the nodes, reflects the network’s overall tendencies. The curl flow, induced by interactions among triangles, reveals rotational behaviors within the network. Meanwhile, the harmonic flow identifies closed loop structures and uncovers the underlying topological features of the temporal network.

The gradient and curl components of the edge flow are obtained via least-squares minimization as projections onto the respective orthogonal planes (curl and harmonic):

\[
    \bar{\bm{\Theta}}_G^{\text{FAST}}(t) = \arg \min_{s(t)} \| \bar{\bm{\Theta}}^{\text{FAST}}(t) - \delta_0 s(t) \|^2,
\]

\[
    \bar{\bm{\Theta}}_C^{\text{FAST}}(t) = \arg \min_{\varphi(t)} \| \bar{\bm{\Theta}}^{\text{FAST}}(t) - \delta_1^\top \varphi(t) \|^2,
\]

with the harmonic component calculated as the residual:

\[
    \bar{\bm{\Theta}}_H^{\text{FAST}}(t) = \bar{\bm{\Theta}}^{\text{FAST}}(t) - (\bar{\bm{\Theta}}_G^{\text{FAST}}(t) + \bar{\bm{\Theta}}_C^{\text{FAST}}(t)).
\]

\begin{figure}[ht]
\centering
\includegraphics[width=\columnwidth]{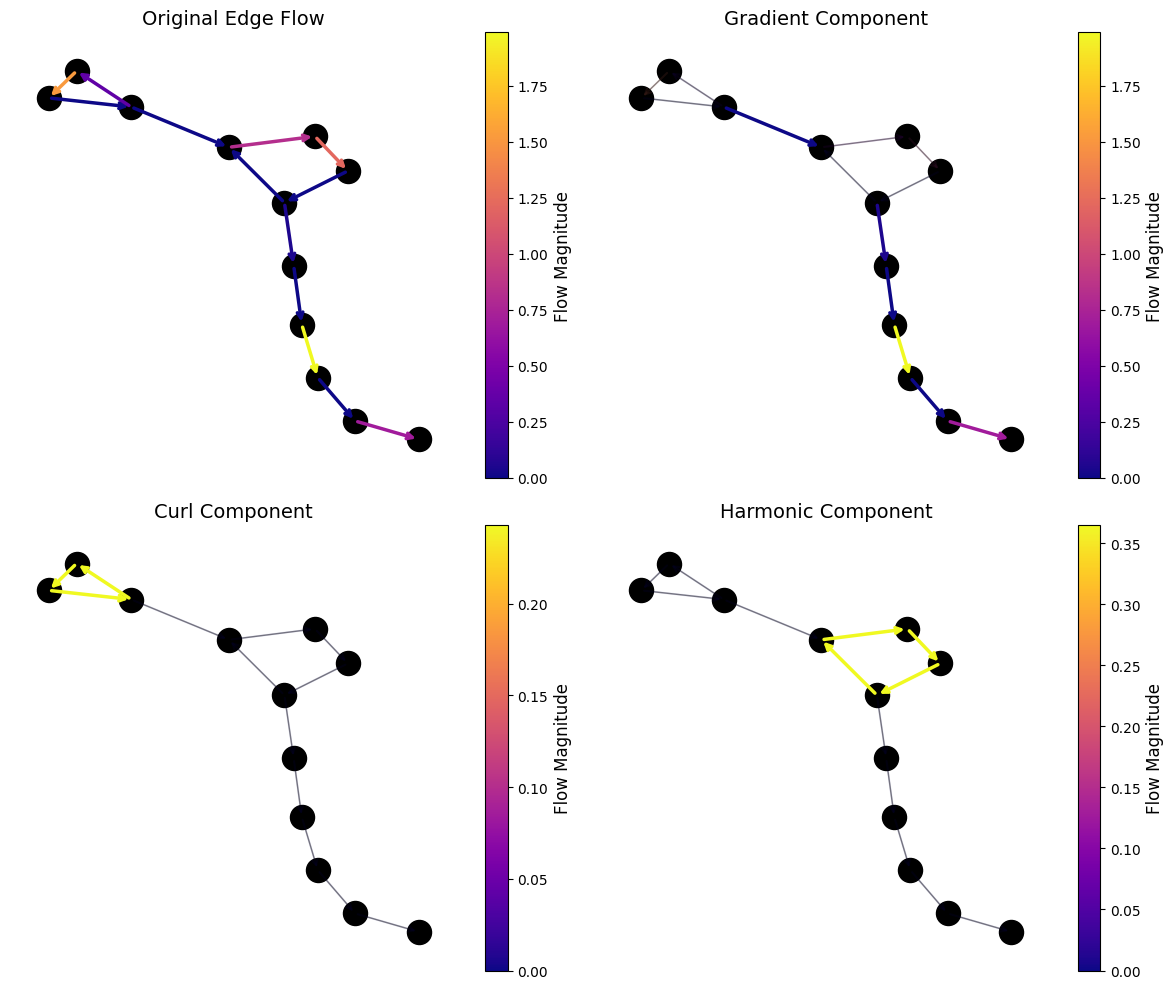}
\caption{A simplified visual representation of the Hodge decomposition of edge flow into its fundamental components: gradient, curl, and harmonic flow. Here chains correspond to gradient flow, triangles represent curl flows and loops characterize harmonic flow.}
\label{fig:example_figure}
\end{figure}

\subsection{Data}

This study examines two datasets of patients with mild cognitive impairment (MCI) due to Alzheimer’s Disease (AD) \cite{mario}, focusing on the pre-dementia stage. The datasets are distinguished in considering familial (MCI-FAM) and sporadic (MCI-SPO) \cite{pietto,hc} groups. MCI-FAM patients carry a genetic mutation and are at high risk for early-onset AD \cite{mcifam}, while MCI-SPO patients have an undetermined risk. Both groups are compared to control participants without genetic mutations or psychiatric/neurological disorders.

The data consists of EEG recordings: 128-channel EEG data recorded at 512 Hz using a Biosemi Active Two System for MCI-SPO patients and 60-channel EEG data recorded at 500 Hz using a Neuroscan SynAmps 2.5 system for MCI-FAM patients\cite{Roy2024}. The EEG signals were band-pass filtered from 1 to 100 Hz and down-sampled to 256 Hz.

The VSTM task used in this analysis is the binding task, where participants are presented with arrays of three distinct shapes, each with a unique color. The task involves three phases: encoding, a short delay, and a test period. Participants determine if the objects in two arrays are identical or different. The task is designed to prevent reliance on spatial cues by randomizing object positions and selecting shapes and colors from predefined sets \cite{mario1,mario4,mario7}. Both MCI-FAM and MCI-SPO patients show clinically significant impairments in the binding task compared to healthy controls, with MCI-FAM performing worse clinically in the task\cite{mario7,mario2}.

Signal pre-processing was applied to the EEG data of the encoding phase, filtering it into Delta (0-4Hz), Theta (4-8Hz), Alpha (8-12Hz) and Beta (12-16Hz) Frequency bands with each epoch lasting 1 second post stimuli exposure.

\section{Results}

For each temporal window we take the mean of the vectorized connectivity profile for each orthogonal component. This gives us a single value for each participant at each time window. 
We perform the Wilcoxon Rank-Sum test for significant differences as a non-parametric approach to detect differences at the group level between controls and patients for each orthogonal component. We thus get 3 sets of $p$-values corresponding to each component at each window. False Discovery Rate Correction is applied to account for false positives, however, this approach inherits the noise resistance of the FAST filter and thus we expect fewer false positives \cite{Roy2024}. Cohen's $d$ effect sizes are also computed with a negative value indicating an increased effect in the patient group.

\subsection{MCI-FAM}

\begin{table}[ht]
\centering
\caption{MCI-FAM Results. Node Function: Local Dirichlet Energy, Sparsity: Top 5\% Connections, Number of Windows: 10.}
\resizebox{\columnwidth}{!}{%
\begin{tabular}{@{}llllll@{}}
\toprule
\textbf{Freq. Band} & \textbf{Time Window}  & \textbf{P-Value} & \textbf{FDR P-Value} & \textbf{Effect Size} & \textbf{Component} \\ \midrule
Delta               & \textbf{0.8-0.9}      & 0.0022           & 0.0220              & -1.4682             & Curl               \\
Delta               & 0.9-1.0              & 0.0113           & 0.0566              & -1.4511             & Curl               \\ \midrule
Theta               & \textbf{0.8-0.9}      & 0.0006           & 0.0058              & -1.6955             & Curl               \\
Theta               & 0.4-0.5              & 0.0312           & 0.3121              & -1.1271             & Harmonic           \\ \midrule
Alpha               & 0-0.1                & 0.0140           & 0.1402              & 1.0807              & Gradient           \\
Alpha               & 0.4-0.5              & 0.0376           & 0.1505              & 1.1777              & Gradient           \\
Alpha               & 0.5-0.6              & 0.0452           & 0.1505              & 1.1570              & Gradient           \\ \midrule
Beta                & 0-0.1                & 0.0452           & 0.4516              & 0.8829              & Curl               \\
Beta                & 0-0.1                & 0.0312           & 0.3121              & -0.9490             & Harmonic           \\
Beta                & 0.3-0.4              & 0.0173           & 0.0863              & 1.0494              & Gradient           \\
Beta                & 0.4-0.5              & 0.0058           & 0.0580              & 1.2730              & Gradient           \\ \bottomrule
\end{tabular}%
}
\end{table}

Table 1 presents significant $p$-values at the 5 percent level and after FDR correction for each frequency band. Most differences are concentrated in the late and middle ranges. Highly significant results in the delta and theta bands, passing FDR correction in the 0.8-0.9 post-stimulus window, link to the Late Positive Potential (LPP), an event-related potential (ERP) \cite{agustin} associated with explicit recognition memory and the binding task \cite{pietto,erp}. Both bands show increased values in the patient group. Using the local Dirichlet energy (equation (2)), this indicates higher local variability in MCI-FAM patients at key global connections for the binding task. This result, associated with the curl component, suggests greater dysregulation or variability in interactions among triangles in the connectivity profile.

Significant differences are also seen in the 0.4-0.6 range across theta, alpha, and beta bands, consistent with the P300 range, an ERP previously linked to this group \cite{Roy2024,p3,mario3}. In the theta band, the harmonic component reveals these differences with larger values in patients, indicating increased local variability in loop-like structures. In contrast, the gradient component in the alpha and beta bands shows P300 differences as reduced local variability in patients.

The alpha and beta bands exhibit significant differences in the early 0-0.1 window, indicating the P100 ERP \cite{erp}. In the alpha band, the gradient component shows lower local variability in controls, while in the beta band, the curl and harmonic components show lower and higher variability in patients, respectively. The P100 has also been highlighted in prior dynamic functional connectivity analyses, particularly in the alpha band \cite{mde}.

\begin{figure}
\centering
\includegraphics[width=\columnwidth]{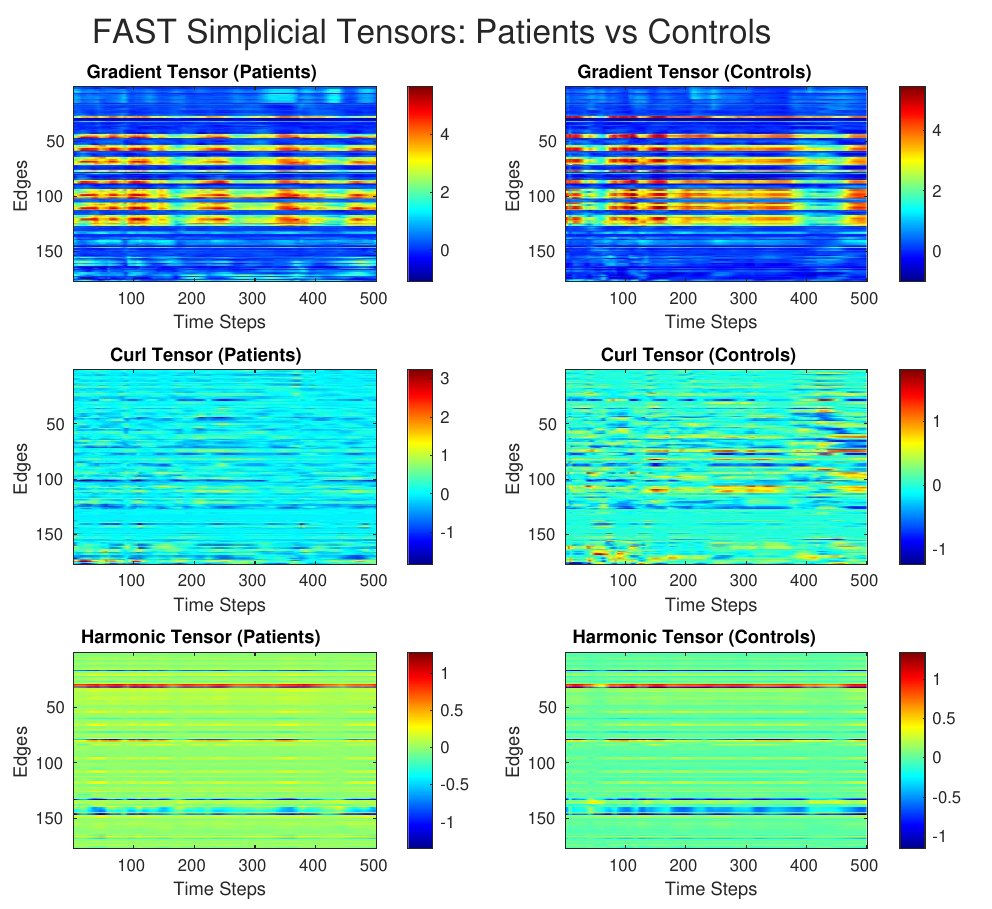}
\caption{Average edge flows for the top 10 \% of the strongest global connections across different orthogonal planes at sample resolution, measured during a time-locked VSTM Binding Task in the Theta Band, for Familial Alzheimer's Disease (MCI-FAM) patients and healthy controls}
\label{fig:example_figure}
\end{figure}

\subsection{MCI-SPO}
\begin{table}[ht]
\centering
\caption{MCI-SPO Results. Node Function: Instantaneous Correlation, Sparsity: Top 1\% Connections, Number of Windows: 15.}
\resizebox{\columnwidth}{!}{%
\begin{tabular}{@{}llllll@{}}
\toprule
\textbf{Freq. Band} & \textbf{Time Window}  & \textbf{P-Value} & \textbf{FDR P-Value} & \textbf{Effect Size} & \textbf{Component} \\ \midrule
Delta               & 0.8-0.87             & 0.0460           & 0.6903              & 0.8579             & Gradient           \\
Delta               & 0.73-0.8             & 0.0383           & 0.1652              & 0.8336             & Harmonic           \\
Delta               & \textbf{0.8-0.87}    & 0.0013           & 0.0190              & 1.3322             & Harmonic           \\
Delta               & 0.87-0.93            & 0.0261           & 0.1652              & 0.9660             & Harmonic           \\ \midrule
Theta               & \textbf{0.33-0.4}    & 0.0017           & 0.0248              & 1.4307             & Gradient           \\
Theta               & 0.33-0.4             & 0.0057           & 0.0860              & 1.1322             & Curl               \\
Theta               & 0.47-0.53            & 0.0261           & 0.3908              & -0.9978              & Harmonic           \\ \midrule
Alpha               & 0.93-1.0             & 0.0348           & 0.4507              & 0.7973             & Curl               \\ \bottomrule
\end{tabular}%
}
\end{table}

Table 2 represents the results for the MCI-SPO dataset, we use instantaneous correlation as the node function, representing synchronization of connectivity. The delta band shows a significant result passing FDR correction in the 0.8–0.87 window, overlapping with the LPP range observed in the MCI-FAM group. However, this time, the harmonic component is primarily implicated, with patients showing significantly reduced connectivity in loop-like structures. Additionally, the alpha band is involved in this LPP response, with reduced connectivity in triangular flows in patients.

The P300 is also implicated, particularly in the theta band during the early 0.33–0.4 window (P3a), as shown by the gradient, harmonic, and curl components. Patients exhibit reduced connectivity in the gradient and curl components but increased connectivity in the harmonic component during the later 0.47–0.53 window (P3b). This delayed increase likely reflects a delay in the P300 response, previously reported in Alzheimer's disease-related impairments \cite{delay}. Notably, this delayed response is most pronounced in loop-like connectivity structures.

\section{Discussion}
In this work, we have developed a novel framework for the analysis of Dynamic Functional Connectivity in EEG signals using tools from topological data analysis and network science. This allowed us to go beyond pairwise interactions in transient connectivity profiles and consider higher-order interactions in the networks.

While higher-order interactions have long been hypothesized to be useful in the analysis of neurological systems, computational limitations and a lack of interpretability have made them impractical in practice. By employing the FAST filter as a thresholding approach, we can uniformly identify the most relevant global connections across participants. This method significantly accelerates the computation of higher-order structures while retaining high temporal resolution.

Applying our framework to the analysis of Familial and Sporadic Alzheimer-related MCI patients, we uncovered significant differences that align with existing literature and could serve as potential biomarkers for early Alzheimer's detection. These findings were consistent across independent patient groups and experimental setups, reinforcing the robustness of our approach. Additionally, our framework simultaneously detects previously implicated results in this data, providing compelling evidence that higher-order interactions influence transient brain networks. For instance, while both MCI-FAM and MCI-SPO are implicated by the LPP, the specific components differ: MCI-FAM shows implications in the curl component, whereas MCI-SPO involves the harmonic and gradient components.

To our knowledge, this work is the first to extend beyond static networks, enabling efficient analysis of higher-order interactions in Dynamic Functional Connectivity of EEG signals. It effectively handles noise in small temporal windows while representing higher-dimensional structures. We believe this framework opens new possibilities for applying TDA to EEG signals and uncovering biomarkers for neurological disorders that pairwise interactions alone cannot detect. We encourage further exploration in this direction.

\section{Conclusions}

In this work we have introduced the first framework capable of detecting higher order dynamic interactions in EEG networks with a low computational cost. Applied to two independent datasets of Patients with Alzheimer's related MCI undergoing a time-locked VSTM task, we see reproducible differences across both datasets that align strongly with previous EEG studies on the task and thus may be used as potential biomarkers. We show how these differences are implicated by specific components each with an independent geometric flow revealing specific insights into how neurological disorders disrupt these networks. We have opened up the possibility for more applications of higher order interactions to dynamic brain networks and encourage work in this domain extending beyond just the EEG.

\end{document}